\documentclass[12pt]{article}

\usepackage{graphics,epsfig}

\def\dalemb#1#2{{\vbox{\hrule height .#2pt
        \hbox{\vrule width.#2pt height#1pt \kern#1pt
                \vrule width.#2pt}
        \hrule height.#2pt}}}

\let\a=\alpha \let\b=\beta  \let\d=\delta \let\e=\epsilon
\let\z=\zeta  \let\th=\theta  
\let\l=\lambda \let\m=\mu  \let\x=\xi  

        \let\Th=\Theta 
\let\X=\Xi  \let\S=\Sigma  \let\Y=\Psi
 
\let\la=\label  
  
\def\nn{\nonumber} \def\bd{\begin{document}} \def\ed{\end{document}}
\def\ds{\documentstyle} \let\fr=\frac \let\bl=\bigl \let\br=\bigr
\let\Br=\Bigr \let\Bl=\Bigl
\let\bm=\bibitem
\let\na=\nabla
\def\tU{{\widetilde U}}
\let\pa=\partial \let\ov=\overline
\def\ie{{\it i.e.\ }}
\newcommand{\be}{\begin{equation}}
\newcommand{\ee}{\end{equation}}
\def\ba{\begin{array}}
\def\ea{\end{array}}
\def\ft#1#2{{\textstyle{{\scriptstyle #1}\over {\scriptstyle #2}}}}
\def\fft#1#2{{#1 \over #2}}
\def\F#1#2{{ F_{#1}^{(#2)} }}
\def\cF#1#2{{ {\cal F}_{#1}^{(#2)} }}

\def\R{{\bf R}}
\def\sst#1{{\scriptscriptstyle #1}}
\def\oneone{\rlap 1\mkern4mu{\rm l}}
\def\e7{E_{7(+7)}}
\def\td{\tilde}
\def\wtd{\widetilde}
\def\im{{\rm i}}
\def\bog{Bogomol'nyi\ }
\newcommand{\ho}[1]{$\, ^{#1}$}
\newcommand{\hoch}[1]{$\, ^{#1}$}
\newcommand{\bea}{\begin{eqnarray}}
\newcommand{\eea}{\end{eqnarray}}
\newcommand{\ra}{\rightarrow}
\newcommand{\lra}{\longrightarrow}
\newcommand{\Lra}{\Leftrightarrow}
\newcommand{\ap}{\alpha^\prime}
\newcommand{\bp}{\tilde \beta^\prime}
\newcommand{\cB}{{\cal B}}
\newcommand{\cO}{{\cal O}}
\newcommand{\vecx}{\vec{x}}
\newcommand{\vecy}{\vec{y}}
\newcommand{\vecp}{\vec{p}}
\newcommand{\vecq}{\vec{q}}
\newcommand{\tr}{{\rm tr} }
\newcommand{\Tr}{{\rm Tr} }
\newcommand{\NP}{Nucl. Phys. }

\newcommand{\cL}{{\cal L}}
\newcommand{\cA}{{\cal A}}
\newcommand{\cD}{{\cal D}}

\def\sst#1{{\scriptscriptstyle #1}}
\def\0{{\sst{(0)}}}
\def\1{{\sst{(1)}}}
\def\2{{\sst{(2)}}}
\def\3{{\sst{(3)}}}
\def\4{{\sst{(4)}}}
\def\5{{\sst{(5)}}}
\def\6{{\sst{(6)}}}
\def\7{{\sst{(7)}}}
\def\8{{\sst{(8)}}}
\def\ve{\varepsilon}
\def\vf{\varphi}
\def\F{\Phi}
\def\wg{\wedge}

\newcommand{\tamphys}{\it 
}

\newcommand{\auth}{AUTHORS}

\def\thb{\bar{\theta}}
\def\Thb{\bar{\Theta}}
\def\barp{\bar{p}}
\def\barq{\bar{q}}
\def\barc{\bar{c}}
\def\bard{\bar{d}}
\def\e{\epsilon}

\def \bi{\bibitem}
\def \la {\label}

\def \l {\lambda}
\def\foot{\footnote}
\def \tl  {{\tilde \l}}
\def \sql {{\sqrt \l}}
\def \adss {$AdS_5 \times S^5$\ }
\newcommand{\rf}[1]{(\ref{#1})}
\def \ov {\over}

\def\th{\theta}
\def\Th{\Theta}
\def\vth{\vartheta}
\def\btheta{{\bar\theta}}
\def\ttheta{{{\tilde\theta}}}
\def\bttheta{{{\bar\ttheta}}}
\def\vth{\vartheta}

\def\ra{\rightarrow}
\def\N{{\cal N}}
\def\F{{\cal F}}
\def\uM{\underline{M}}
\def\uN{\underline{N}}
\def\uP{\underline{P}}
\def\cc{\circ}
\def\eqv{\equiv}

\def\ni{\noindent}

\def\Ep{E^{{}^{(+)}}}
\def\Em{E^{{}^{(-)}}}

\def\Mp{M^{{}^{(+)}}}
\def\Mm{M^{{}^{(-)}}}

\def \ha{{1\ov 2}}

\def\r{\rho}

\def\Y{{\rm Y}}
\def\X{{\rm X}}
\def\tY{\tilde{\rm Y}}
\def\tX{\tilde{\rm X}}
\def\dY{\dot{\rm Y}}
\def\dX{\dot{\rm X}}

\def \J {\mathcal{J}}
\def \del {\partial}

\def\dF{\dot{F}}
\def\dG{\dot{G}}
\def\df{\dot{f}}
\def \E {{\cal E}}
\def \S {{\cal S}}
\def \J {{\cal J}}

\def\ms{\mathcal{S}}
\def\mj{\mathcal{J}}
\def\soj{\fr{\ms}{\mj}}
\def \R {{\bf R}}
\def \om {\omega}
\def \bE {\bar E}
\def \x {{\cal X}}

\def \bi{\bibitem}
\def \la {\label}

\def \l {\lambda}
\def\foot{\footnote}
\def \tl  {{\tilde \l}}
\def \sql {{\sqrt \l}}
\def \adss {$AdS_5 \times S^5$\ }
\def \ov {\over}

\def \varpi {{\rm w}}

\def\thb{\bar{\theta}}
\def\Thb{\bar{\Theta}}
\def\zb{\bar{z}}
\def\psib{\bar{\psi}}
\def\barp{\bar{p}}
\def\barq{\bar{q}}
\def\barc{\bar{c}}
\def\bard{\bar{d}}
\def\e{\epsilon}
\def\wb{\bar{w}}
\def\lb{\bar{\l}}
\def\mb{\bar{m}}
\def\nb{\bar{n}}
\def\Jb{\bar{J}}
\def\Nb{\bar{N}}

\def\At{\tilde{A}}
\def\Bt{\tilde{B}}
\def\Ct{\tilde{C}}
\def\Dt{\tilde{D}}
\def\Et{\tilde{E}}
\def\Ft{\tilde{F}}
\def\Gt{\tilde{G}}
\def\Mt{\tilde{M}}
\def\at{\tilde{a}}
\def\bt{\tilde{b}}
\def\ct{\tilde{c}}
\def\dt{\tilde{d}}
\def\et{\tilde{e}}
\def\ft{\tilde{f}}
\def\gt{\tilde{g}}
\def\ola{\overleftarrow}
\def\ora{\overrightarrow}
\def\at{\tilde{\a}}

\def\ps{\rlap{\, /}\;\,p }
\def\ks{\rlap{\, /}\;\,k }

\def\gym{g_{YM}}

\def\adot{\dot{a}}
\def\bdot{\dot{b}}
\def\bpa{\bar{\pa}}

\begin{document}
\overfullrule=0pt
\parskip=2pt
\parindent=12pt
\headheight=0in \headsep=0in \topmargin=0in
\oddsidemargin=0in

\vspace{ -3cm}
\thispagestyle{empty}

\begin{center}

{\Large\bf Vertex operator formulation of scattering around
black-hole
  }

 \vspace{.5cm} { I.Y. Park  }\\
 \vskip 0.2cm

{\it Kavli Institute for Theoretical Physics,\\
 Santa Barbara, California, USA \\
}
 \vspace{0.8cm}
{\it Center for Quantum Spacetime, Sogang University\\
Shinsu-dong 1, Mapo-gu, 121-742, South Korea \\
}

 \vspace{0.4cm}
 and\\
 \vspace{0.4cm}
{\it Division of Physical and Natural Sciences{\footnote{Home
instutution}},
Philander Smith College\\
Little Rock, AR 72202, USA \\
inyongpark05@gmail.com}


\end{center}

 \vspace{0.1cm}

 \begin{abstract}
We propose a full-fledged open string framework that seems suited to
study the black hole information paradox. We set up a configuration
to compute the scattering amplitude of a IIB open string around a
D5-brane. The D5-brane is situated at the origin of a transverse
D3-brane. A string perturbation theory is employed where the
geometry of the D5-brane is treated as a potential. We reason that
the setup is capable of reconciling the unitary evolution of states
and information loss that is measured by an observer on the D3
brane. With the configurations of these kinds, the information loss
is an apparent phenomenon: it is just a manifestation of the fact
that the D3-observer does not have access to the ``hair" of the D5
black brane.

\end{abstract}
\newpage

\setcounter{equation}{0}
\setcounter{footnote}{0}
\setcounter{section}{0}



Black holes have played important roles in recent developments in
string theory \cite{Strominger:1996sh,Gubser:1996de}. They appear as
relatively simple solutions of various supergravity theories whose
horizon areas yield Bekenstein-Hawking entropies. (See, for example
\cite{Behrndt:1998eq,Sen:2004dp,Dabholkar:2004dq,Myung:2007xd}.) The
microscopic origin of the entropy was then understood
\cite{Strominger:1996sh,Gubser:1996de,Wadia:2000wy, Peet:2000hn} by
making a connection with D-branes
\cite{Polchinski:1995mt,Polchinski:1996na,Pol,Johnson:2000ch}. One
well-known puzzle in black hole physics is the black hole
information paradox. Relatively recent discussions on the topic can
be found in
\cite{Banerjee:2009uk,Mathur:2008nj_1,Mathur:2008nj_2,Iizuka:2008eb,Horowitz:2009wm}.
 Since string theory
is capable of describing gravitation, it should provide a proper
setup to tackle the problem. There have also been studies on
absorption cross sections of various black holes, e.g.,
\cite{chan,Das:1996we,Emparan:1997iv,Jung:2004yh}. Most of these
works are based on the low energy effective actions of string
theory, supergravity or gauge/DBI theory. What is desired is a
full-fledged vertex operator formulation that is designed, with a
``realistic" touch,
 to compute various scattering amplitudes around a black hole.
 In this paper, we propose one such setup.

The system that we consider consists of a D5-brane and a
complementary D3 brane. As shown in the figure below, they are
entirely perpendicular. The scattering open strings move
 on the D3 brane. It also hosts an observer, therefore we may call it
  a ``brane-world". To the observer, the D5 brane appears as a {\em hole}
  instead of a brane. Namely, confined within the brane, the observer
 only sees a ``dot", the black hole. The scattering {\em around} the black hole is
measured. A type IIB supergravity solution that corresponds to such
a configuration have not been established in the literature although
we expect its existence. The configuration might be obtained from a
known solution. We briefly comment on this later. Intuitively it is
obvious that the D5-brane provides an ``escape passageway" for some
of the incoming strings. This will cause a loss in the flux that is
measured by the D3-observer, which will be interpreted as
information loss. With the setup under consideration (and the
similar kinds), it is clear that the loss is an apparent phenomenon:
it is a simple reflection of the D3 observer's inaccessibility to
the ``hair" of the D5 brane. One of the goals of this paper is to
provide a quantitative description of the scattering.

The picture can be put on computational ground with the following
philosophy. What we propose as part of the setup is two-sided
concerning how to deal with each of the geometries, the geometry due
to D5 and that of the D3. It is based on our view that the two
geometries should have different interpretations and therefore play
different roles. In the spirit of previous works
\cite{Park:2007mc,Park:2008sg,Park:2008fp,Park:2009ki} we attribute
the geometry of D5-brane to the loop effects of the open strings
that live in the D5-brane. Such physics is only
 indirectly visible, and only in a limited manner, to an observer who lives
 on the D3. Most of the detailed physics will be hidden: we propose
 to treat the geometry of the D5 brane as a {\em potential} that is felt by the
scattering open strings. What it means in terms of the action is
that we start with a non-linear sigma model
\cite{Cvetic:1999zs,Sahakian:2004gy,Mizoguchi:2002qy} of D5-brane
geometry and view it as consisting of free pieces plus the potential
terms. The potential will then affect the dynamics of the open
strings moving on the D3. It is the part of the proposal that makes
the setup different from those in the other literature. The second
ingredient-which is also in the spirit of the previous works- is
that we associate the geometry of the D3 brane with the loop
effects\footnote{The connection between loop effects and geometry
goes back to the Fischler-Susskind mechanism \cite{Fischler:1986ci}.
The connection between open string loop effects and the D-brane
geometry in the current context has been established at one-loop in
\cite{Park:2008fp} and partially established at two-loop in
\cite{Park:2009ki}. Work is in progress \cite{progress} for the
three-loop order.} of the scattering open string on the D3-brane. It
implies, in the case where one considers only the tree level
scattering (as we do in this work), that one may disregard the
curved geometry due to the D3-brane: it will only play a role as
counter terms in loop scattering diagrams.

The discussion so far leads to the following overall picture: we
consider an open string moving on the D3-brane whose vertex operator
has been constructed in \cite{Park:2007mc} following \cite{gsw}. As
in the standard quantum mechanical formulation of scattering, we
consider a two-point amplitude. The effect of the presence of a
black hole in general is then introduced as a vertex operator
insertion in the correlation function. The precise form of the
potential is determined based on the two-dimensional non-linear
sigma model of the D5-brane. The correlator can be evaluated by
following the method in the previous works. Towards the end, we note
that, in a full string description, genuine information loss does
not occur in a generic configuration of the present kind.\footnote{
There is a difference between an open string and a closed string. We
will comment
 on this as well.}  \\

\begin{figure}[!ht]
\centerline{
     \hspace{1in}   \begin{minipage}[b]{20cm}
               \epsfxsize=16cm
                \epsfbox{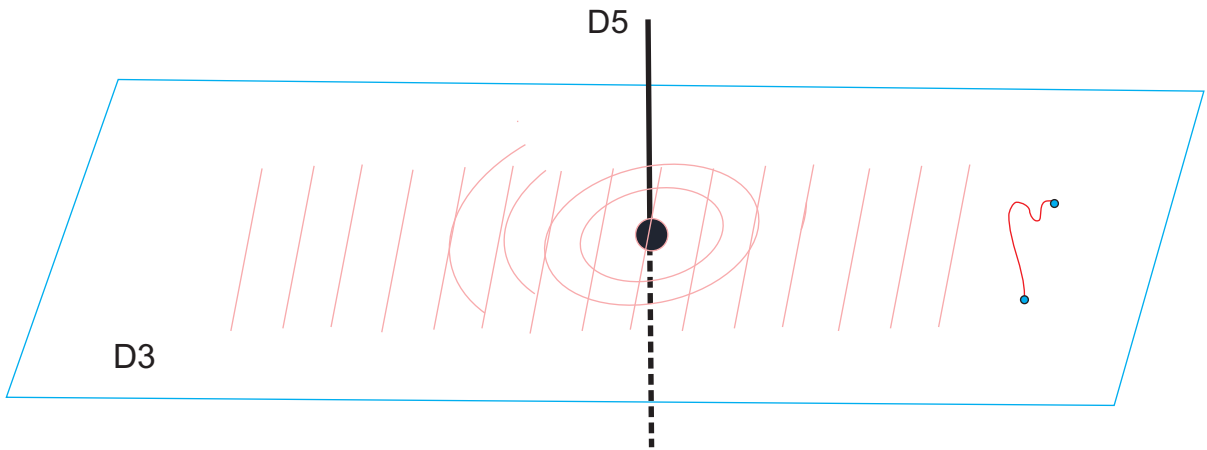}
        \end{minipage}
} \caption{An open string scattering around a D5-brane} \label{fig1}
\end{figure}

The detailed computation will be taken below but it
 may be worth thinking about what one should expect as a result
 of the computation. In order to answer the question, we try to develop a further
intuition on the system. The first issue that we would like to
address is a potential obstacle: whether we have a necessary tool
for the computation. In general, studying physics near a black hole
may require a non-perturbative tool that is unavailable. However,
since we are interested mainly in the flux loss, we can adopt a
procedure that is similar to the one developed in
\cite{Park:2008fp}. In this procedure, we consider what we call a
large $R_0$-expansion where one expands the spatial D3-brane
coordinate as
 \bea
 \mbox{X}^u=<X^u>+X^u=X_0^u+X^u,\quad R_0^2=\sum_u (X_0^u)^2
 \label{shiftedX}
 \eea
Invoking Ehrenfest theorem, we interpret $R_0$ as the average
location or the location in the classical sense of the
string.\footnote{It is an "average distance" from the black hole and
the location of the string.}
 In
other words, by considering the large-$R_0$ expansion we are
analyzing the dynamics of a string which has an average location at
$R_0$.\footnote{One may pause and wonder whether such a large
distance expansion can properly capture the black hole information
physics. With regard to this issue, we focus on whether there is a
Hawking radiation. If there is, it must be also observed from far
away from the black hole. The large distance expansion must be
adequate for that.}

We are ready to discuss what to expect as an outcome of the
computation. For that purpose, we use an analogy with the quantum
mechanical formulation of scattering. To avoid unnecessary
complications, we consider non-relativistic scattering. The
discussion will be heuristic.

If one sets out to analyze non-relativistic scattering around the
D5-brane, one will consider a wave-function which schematically
takes the form of
 \bea
 e^{ik\cdot x}+\fr{e^{ikr}}{r^n}f(k',k)+\b({\bf r})\psi_{D5}
 \label{3superposition}
 \eea
The first two terms are standard except that the power of
$\fr{1}{r}$ in the second term has been set to
 $n$ with $n>1$. In a flux-conserving scattering which is usually the case
 in quantum mechanics, the power is $n=1$. This will not be the case for
 the system under consideration: the D3-observer will measure flux loss since part
of the flux will be absorbed into the D5-brane. The absorbed part
will form a wave-function that is represented by the third term. The
profiling function $\b({\bf r})$ should be sharply peaked at the
three dimensional origin ${\bf r}=0$.
 The third term will contribute to the large-$R_0$ scattering
 only through making $n$ higher than one. From the setup, the following
 property of the cross section
 can be deduced: a cross section in general reveals certain
 properties of the target. It is
based on what/how much comes out of what has entered. The flux loss
in the present setup indicates that there is decay in the wave
function. The (differential) cross section will depend on where the
measurement is being made. In particular, it will be a decreasing
function of $R_0$. There will be
more discussions using (\ref{3superposition}) near the end.\\

There is a large number of literature on scattering that involves
D-branes, \cite{Hashimoto:1996bf,gm,Balasubramanian:1996}, to name a
few. We consider a complementary system of D3/D5. One difference of
this setup is that the observer is confined within the D3-brane. The
precise super-gravity solution has not been obtained. However, there
is a configuration made out of intersecting branes
\cite{Khuri:1993ii,Gauntlett:1996pb,Argurio:1997nh,Ohta:1997gw,
Edelstein:1998vs,Edelstein:2004tp,Cha:2003ws} that is close.

 As
stated in the introduction, two geometries, one from the D5 brane
and the other from the D3 brane, have different meanings. This is
the rationale for different treatments of the two geometries: we
will treat the D5-geometry as a potential whereas the D3-geometry
will be dealt with as the loop effects of the scattering strings.
Since we are only concerned with the tree level scattering, we
disregard the D3-geometry curved terms.

The D5-brane that we consider is given by\footnote{The symbol, $r$,
is transverse to the D5-brane. It is not to be confused with $r$ of
the previous works which is transverse to the D3-brane. For the
``average" location, we are using $R_0$ instead of $r_0$. Since the
roles of the two geometries are very different, so are the roles of
$r_0$ and $R_0$.  }
 \bea
 e^{-\phi}&=&H^{1/2} \nn\\
 ds^2 &=& H^{-\fr14}(dX^m)^2+ H^{\fr34}(dX^\m)^2 \nn\\
 {\cal F}_{ty^1\cdots y^5 \m}&=& -2Q{H^{-2}}\fr{X^\m}{r^4}\nn\\
 H&=& 1+\fr{Q}{r^2},\quad Q= gN\a' \label{D5}
 \eea
The coordinate $X^m$ spans the six Euclidean space and $X^\m$ the
four dimensional space-time. For the scattering amplitude, our
proposal amounts to computing an open string two-point function with
insertion of a potential vertex operator,\footnote{As far as we can
see, this formula is pointed to by an analogy with the standard
quantum mechanical treatment of scattering. If preferred, it may be
taken as part of our proposal.}
 \bea
 <V(x_1,k_1)\;U_{pot}(y)V(x_2,k_2)>
 \label{vuv2}
 \eea
 The locations should be determined based on the physics of scattering.
 A natural choice is $x_1=\infty, x_2=0, y=1$. The precise form of $U_{pot}$
 is determined by considering the
non-linear model of the D5-brane geometry. In the leading order in
$r$, there are only two terms
 \bea
 -\fr12
    \left(\pa_i X^m \pa^i X^n\eta_{mn} H^{-\fr14}
    +\pa_i X^{u} \pa^i X^{v}\eta_{uv} H^{\fr34} \right)
 \eea
 The leading term\footnote{The non-linear sigma model of a D5 contains
 many terms including the parts coming from the Ramond-Ramond fields. They are all
 sub-leading in the large $R_0$-expansion.} in the $R_0$-expansion of this gives
 \bea
 U_{pot}\equiv  \;
    \left(\fr18\fr{Q}{R_0^2}\;\pa_i X^m \pa^i X^n\eta_{mn}
    -\fr38\fr{Q}{R_0^2}\pa_i X^{u} \pa^i X^{v}\eta_{uv}  \right)
    \label{leadingpot}
 \eea
 where $Q$ is a constant in (\ref{D5}). $R_0$ has appeared as a result of
 shifting $\mbox{X}^m$ according to (\ref{shiftedX}) and carrying out the
 large-$R_0$ expansion.
Let's consider the vector boson vertex operator on the D3-brane,
 \bea
 V_{v}(\z,k)&=&
 \z^u (\dot{X}^u-R^{uv}k^v) e^{ik\cdot X}
 \eea
Including the polarization vectors, $(\z^1,\;\z^2)$, and noting that
the first term in (\ref{leadingpot}) do not contribute due to
dimensional regularization, the two-point correlator becomes
 \bea
 &&<V_{v}(x_1,\z^1,k^1)\;U_{pot}(1)\;V_{v}(0,\z^2,k^2)\;>\nn\\
 =&& <V_{v}(x_1,\z^1,k^1)\left[-\fr38\fr{Q}{R_0^2}\;\pa_i X^{u} \pa_j X^{v}\eta_{uv}
    \right]V_{v}(0,\z^2,k^2)\;> \nn\\
 \equiv && \d^{4}(k_1+k_2)I
     \label{2ptkappa}
 \eea
The fermionic terms in $V_{v}(x_1,\z^1,k^1),V_{v}(0,\z^2,k^2)$ do
not contribute for the same reason. Choosing $x_1=\infty, x_2=0,
y=1$ and taking the string measure into account, one gets, after
some algebra which is a simple version of computations in
\cite{Park:2008fp,Park:2009ki},
 \bea
  I&=& \;-\fr{3Q}{4R_0^2}\; \zeta _1.\zeta_2 \label{result} \nn\\
 \eea
The result implies that the cross section decreases as
$\fr{1}{R_0^2}$ after the spherical shell surface integration taken
into account. In particular, it vanishes at infinity. Physically
this means that there is a total absorption of the flux into the D5 brane.\\

We now ponder on the ramifications of the result (\ref{result}) and
possible future directions. Although we have considered a
D3/D5-brane system, qualitative features will be shared by many
other configurations. Generically the wave function will display a
decaying behavior due to the inverse field potential. At infinity
the cross section vanishes. The outgoing wave does not reach
infinity. This feature does not change even for the known extremal
D-5 brane solution. Still the leading behavior is $\fr{1}{R_0^2}$ as
we discuss below. The result then implies that Hawking radiation
does not occur since it would require a non-decaying wave function
as necessary condition. Although we have investigated an extremal
case, it does not seem that near-extremal cases will be different in
this respect. We may consider what would be a near extremal D5
configuration by first considering a near extremal D1/D5 system and
setting the D1-brane charge to zero. To get a non-decaying wave
function, therefore Hawking radiation, the leading power of the
large-$R_0$ expansion of the potential must be $\fr{1}{R_0}$.
However, the non-extremality does not bring that feature. It is
likely that the desired power will come only when there is a
spherical symmetry not only for the potential but also for the
scattering states. Such symmetry will be possessed by a scalar
vertex operator of a closed string, say, dilaton. In that case,
Hawking radiation can be realized. This case will not lead to the
information paradox either: an observer who has access to a closed
string will have access to the detailed physics of the the D5-brane.
The observer should live in the bulk. Therefore, it will be
understood what is causing the apparent non-unitary evolution.

As just discussed, generically Hawking radiation is not realized.
But with a special design, it may be possible. Let us suppose that
we have such a setup and scrutinize the open string case. This will
help us to develop further intuition on the hair and bring out the
tasks involved for quantitative understanding of the information
``loss". We again resort to the non-relativistic quantum mechanical
picture, (\ref{3superposition}), but now with $n=1$,
 \bea
 e^{ik\cdot x}+\fr{e^{ikr}}{r}f(k',k)+\b({\bf r})\psi_{D5}
 \label{3superpositionq}
 \eea
 The third part is the wave function of the D5-brane. It does
not contribute to the large-$R_0$ scattering. It is a product of a
four dimensional wave-function, $\b({\bf r})$, and a wave function
in the D5-brane, $\psi_{D5}$. The wave-function, $\b({\bf r})$,
should be viewed as an invisible part of the 4D wave-function. The
specific question to study is, ``is it possible to change the first
piece and let the change absorbed by the third piece while keeping
the second piece the same?" That would mean that what comes out may
be independent of what has entered other than through the mass (or
the energy) as we will argue now. Writing the ten dimensional
Hamiltonian as a sum of the D3-brane Hamiltonian, $H_4$ and the
transverse one, $H_6$, schematically the Schrodinger equation reads
 \bea
 (H_4+H_6)\left[e^{ik\cdot x}+\fr{e^{ikr}}{r}f(k',k)+\b({\bf
 r})\psi_{D5}\right]
 =E\left[e^{ik\cdot x}+\fr{e^{ikr}}{r}f(k',k)+\b({\bf
 r})\psi_{D5}\right]
 \eea
where $H_4\b({\bf r})=e_4\b({\bf r})$ and $\b({\bf r})\ra 0$ when ${
r}\ra \infty$. We also assume that $H_6$ is purely kinetic. For
$r\ra \infty$
 \bea
 H_4\left[e^{ik\cdot x}+\fr{e^{ikr}}{r}f(k',k)\right]
 \sim  E\left[e^{ik\cdot x}+\fr{e^{ikr}}{r}f(k',k)\right]
 \label{radial}
 \eea
For $r\sim 0$, the $\b({\bf r})$-containing terms will be
dominating,\footnote{Remember that the $\fr{1}{r}$-behavior of the
second term is only asymptotically valid.}
 \bea
&& \psi_{D5} H_4\b({\bf r})+\b({\bf r})H_6\left[\psi_{D5}\right]
 \sim E\left[\b({\bf r})\psi_{D5}\right] \nn\\
\Rightarrow && H_6 \psi_{D5} \sim (E-e_4)\psi_{D5} \label{smallr}
 \eea
The above two equations, (\ref{radial}) and (\ref{smallr}), achieve
the goal: the change in the 4D wave function, i.e., the change in
$\b({\bf r})$ gets absorbed by the change in $\psi_{D5}$, therefore
not affecting the radially outgoing wave function.  Let us set
 \bea
 M_{D5}=E-e_4
 \eea
where $M_{D5}$ would be the mass or the energy of the D5-brane. It
shows that what comes out is independent of what has entered other
than through the mass of the D5-brane. So the D3-brane observer will
measure the information loss while 10D physics still preserves the
unitary evolution of states.

One may pursue the following future directions: the current setup
may be useful to understand the black hole entropy at the full
vertex operator level. One direction is along a near-extremal D5
case. The near-extremal brane geometry should presumably be
associated with the excitations of massive open strings on the
D5-brane. They must be the hair. It will be interesting to compute
the entropy of the massive open string states and compare it with
the Bekenstein-Hawking entropy. It will also be interesting to study
scattering of a scalar state of a closed string. That will make a
connection
with existing literature.\\

We end by posing a bold question. Based on the discussion so far,
the information paradox does not seem to occur in string theory,
regardless of whether there is Hawking radiation or not. However, it
 occurs in theories based on four dimensions: should the black hole
information paradox based on four dimensional models be taken as an
indication that our world has more dimensions
 than four?

\vspace{1in} \ni {\bf Acknowledgements:}

\ni I acknowledge the KITP scholar program of UCSB. This research
was supported in part by the National Science Foundation under Grant
No. PHY05-51164. I thank Bumhoon Lee for his hospitality during my
stay at CQUeST (Sogang university) where the work was completed.
This work was supported in part by Korea Science and Engineering
foundation through CQUeST with grant number R11-2005-021.



\end{document}